\documentclass[aps,showpacs,tightenlines,twocolumn,nofootinbib,nobibnotes,superscriptaddress]{revtex4-1}
\usepackage{amsmath,amssymb,amsfonts,bm}
\usepackage{graphicx}
\usepackage{epstopdf}
\usepackage{dcolumn}
\usepackage{mathrsfs}
\usepackage{tgtermes}
\usepackage[colorlinks=true,linkcolor=red,citecolor=blue, urlcolor=blue,bookmarks=false]{hyperref}
\bibpunct{[}{]}{,}{n}{}{}
\usepackage{verbatim}

\begin{document}
\title{Disorder-induced chiral and helical Majorana edge modes in a two-dimensional Ammann-Beenker quasicrystal}
\date{\today }
\author{Chun-Bo Hua}
\affiliation{Department of Physics, Hubei University, Wuhan 430062, China}
\author{Zheng-Rong Liu}
\affiliation{Department of Physics, Hubei University, Wuhan 430062, China}
\author{Tan Peng}
\affiliation{Department of Physics, Hubei University, Wuhan 430062, China}
\author{Rui Chen}
\affiliation{Shenzhen Institute for Quantum Science and Engineering and Department of Physics, Southern University of Science and Technology, Shenzhen 518055, China}
\author{Dong-Hui Xu}
\affiliation{Department of Physics, Hubei University, Wuhan 430062, China}
\author{Bin Zhou}\email{binzhou@hubu.edu.cn}
\affiliation{Department of Physics, Hubei University, Wuhan 430062, China}

\begin{abstract}
Recent research on disorder effects in topological phases in quasicrystalline systems has received much attention. In this work, by numerically computing the (spin) Bott index and the thermal conductance, we reveal the effects of disorder on a class D chiral topological superconductor and a class DIII time-reversal-invariant topological superconductor in a two-dimensional Ammann-Beenker tiling quasicrystalline lattice. We demonstrate that both the topologically protected chiral and helical Majorana edge modes are robust against weak disorder in the quasicrystalline lattice. More fascinating is the discovery of disorder-induced topologically nontrivial phases exhibiting chiral and helical Majorana edge modes in class D and DIII topological superconductor systems, respectively. Our findings open the door for the research on disorder-induced Majorana edge modes in quasicrystalline systems.

\end{abstract}

\maketitle

\section{Introduction}
The topological superconductor (TSC), which holds Majorana fermions  \cite{Wilczek2009NP,Service2011Science,Alicea2012RPP,Beenakker2013ARCMP,Elliott2015RMP} and provides a platform for topological quantum computation \cite{Kitaev2003,Nayak2008RMP,Akhmerov2010PRB,Sarma2015Majorana,Zhang2018TQC}, is one of the recent important and highly explored research topics in condensed-matter physics \cite{Read2000PRB,Kitaev2001,Qi2011RMP,Leijnse2012SST,Stanescu2013JPCM,Sato2017RPP,He2018CSB}. And the TSC, a fermionic system described by the fully gapped bulk Hamiltonian, can be classified into the tenfold Altland-Zirnbauer (AZ) symmetry classes based on the three fundamental symmetries containing particle-hole symmetry (PHS), time-reversal symmetry (TRS), and chiral symmetry \cite{AZ1997PRB,Schnyder2009AIP,Kitaev2009AIP,Ryu2010NJP,Chiu2016RMP}. For instance,  two-dimensional (2D) TSCs can be classified into two different categories, including the chiral TSC and time-reversal-invariant (TRI) TSC, according to the three fundamental symmetries. The chiral TSC \cite{Read2000PRB,PhysRevB.82.184516}, which possesses only PHS, is classified into class D of the AZ tenfold classification table and characterized by the $\mathbb{Z}$ topological index (such as the integer Chern number). And the chiral TSC, the superconductor analog of the quantum anomalous Hall insulator, holds chiral Majorana edge modes (MEMs) at its boundary, which is guaranteed by the bulk topological invariant. This is the manifestation of the bulk-boundary correspondence, a guiding principle in the topological phase of matter. Another well-known example is the TRI TSC \cite{PhysRevLett.102.187001} in class DIII of the AZ tenfold classification table, in which the system possesses PHS, TRS, and chiral symmetry. The TRI TSC, a superconductor analog of the quantum spin Hall insulator, is characterized by the $\mathbb{Z}_2$ topological index (such as the spin Chern number). At the boundary of the TRI TSC, the helical MEMs emerge, which are a Kramers pair of time-reversal-related chiral MEMs.

Recently, seeking the MEMs in TSCs has received more attention. Actually, the realization of natural TSCs remains a decade-old outstanding question. Fortunately, the discovery of topological insulators provides a good platform for searching TSCs. In 2008, Fu and Kane proposed that a strong topological insulator proximity coupled with an $s$-wave superconductor can be used to realize TSCs \cite{PhysRevLett.100.096407}. Therefore, an implementation scheme of the chiral MEMs is a hybrid system consisting of a quantum anomalous Hall insulator and an $s$-wave superconductor \cite{PhysRevB.82.184516}. A collection of studies on this scheme has reported theoretical and experimental aspects \cite{PhysRevB.83.100512,PhysRevB.92.064520,PhysRevB.93.161401,PhysRevLett.121.256801,He294Science,PhysRevLett.120.107002,Kayyalha64Science}. Moreover, the helical MEMs in TSCs have also been extensively researched in theory and experiment \cite{PhysRevLett.105.097001,PhysRevLett.111.056402,PhysRevB.83.220510,
PhysRevLett.105.097001,PhysRevLett.108.147003,PhysRevLett.126.137001,Wang104science}, including a typical theoretical proposal for the heterostructure composed of a quantum spin Hall insulator sandwiched by two $s$-wave superconductors with a $\pi$ phase difference \cite{PhysRevB.83.220510} and the experimental signature of helical MEMs revealed in the domain walls of the iron-based superconductor FeSe$_{0.45}$Te$_{0.55}$ \cite{Wang104science}.

Until now, the great majority of studies on TSCs have been implemented in crystalline systems, which can be solved by the topological band theory. It is interesting to note that the TSCs were also recently investigated in quasicrystalline systems \cite{PhysRevLett.116.257002,PhysRevLett.123.196401,PhysRevLett.125.017002,ghadimi2020topological,
PhysRevLett.110.176403,PhysRevB.94.125408,JPSJ2017Fab,PhysRevResearch.3.013148,zeng2020topological,cheng2021fate,PhysRevB.103.104203}, which lack the translational symmetry and cannot be explained by the topological band theory. For example, Fulga \emph{et~al}. \cite{PhysRevLett.116.257002} proposed that the chiral TSC can be realized on an eightfold Ammann-Beenker (AB) tiling quasicrystalline lattice (QL), in which the chiral MEMs are characterized by the nonzero pseudospectrum $\mathbb{Z}$ index \cite{LORING2015383} and the quantized thermal conductance. Furthermore, Ghadimi \emph{et~al.} \cite{ghadimi2020topological} showed that both the fivefold Penrose tiling and the eightfold AB tiling QLs can be used as a platform to achieve the TSC with chiral MEMs, where the bulk topology is characterized by the unity Bott index. But as we know, the investigation of the TRI TSCs with helical MEMs in quasicrystalline systems is still lacking. Besides the TSCs, multiple topological phases of matter have also been proposed for quasicrystalline systems in recent years \cite{PhysRevLett.111.226401,PhysRevB.91.085125,PhysRevB.100.214109,
PhysRevLett.121.126401,PhysRevB.98.125130,PhysRevB.100.115311,PhysRevB.103.085307,PhysRevB.100.085119,
PhysRevLett.124.036803,PhysRevB.102.241102,PhysRevResearch.2.033071,
PhysRevB.101.041103,PhysRevLett.109.106402,PhysRevLett.109.116404,PhysRevB.88.125118,PhysRevX.6.011016,
PhysRevLett.119.215304,PhysRevX.9.021054,PhysRevLett.123.150601,PhysRevB.101.115413,Huang_2020,
PhysRevLett.122.237601,PhysRevB.100.054301,PhysRevB.102.024205,PhysRevB.101.125418,PhysRevB.101.020201,PhysRevA.103.033325}, such as quantum Hall insulators
\cite{PhysRevLett.111.226401,PhysRevB.91.085125,PhysRevB.100.214109}, quantum spin Hall insulators \cite{PhysRevLett.121.126401,PhysRevB.98.125130,PhysRevB.100.115311,PhysRevB.103.085307,PhysRevB.100.085119}, and higher-order topological insulators \cite{PhysRevLett.124.036803,PhysRevB.102.241102,PhysRevResearch.2.033071}. Experimentally, the photonic quasicrystals \cite{PhysRevLett.110.076403} and the quasiperiodic acoustic waveguides \cite{PhysRevLett.122.095501} can be employed as platforms to realize the topological phase of matter in QLs.

In addition, one of the most significant properties of the topological phases of matter is the robustness of the edge states against weak disorder, which is protected by the bulk topology. When the energy gap is closed by strong disorder, a topological phase transition appears, and the topology disappears. The more intriguing finding is that disorder can encourage the emergence of a topologically nontrivial phase in an initially clean and normal system \cite{PhysRevB.100.115311,PhysRevB.103.085307,PhysRevLett.102.136806,
PhysRevLett.105.115501,Zhang_2013CPB,PhysRevB.92.085410,PhysRevLett.116.066401,PhysRevB.100.054108,
PhysRevB.80.165316,PhysRevLett.103.196805,PhysRevLett.105.216601,
PhysRevB.84.035110,PhysRevLett.113.046802,PhysRevB.91.214202,PhysRevB.96.205304,Wu_2016CPB,
PhysRevB.93.125133,Qin2016SR,PhysRevB.103.115430,Habibi2018PRB,Lieu2018PRB,PhysRevB.100.205302,
PhysRevLett.125.166801,PhysRevB.103.085408,
PhysRevLett.115.246603,PhysRevB.95.245305,PhysRevB.97.235109,
Zhang_2020SCPMA,luo2019nonhermitian,PhysRevA.101.063612,Liu_2020CPB,liu2021real,
PhysRevB.93.214206,PhysRevLett.114.056801,PhysRevLett.125.217202,PhysRevB.103.224207,hu2021bulkboundary}. Disorder-induced topological phases have been achieved in various experiment platforms \cite{Meier2018Science,Stutzer2018Nature,PhysRevLett.125.133603,PhysRevLett.126.146802}, such as one-dimensional disordered atomic wires \cite{Meier2018Science} and photonic lattices \cite{Stutzer2018Nature}.
The pioneering work is the proposal of topological Anderson insulators in HgTe quantum wells by Li \emph{et~al.} in 2009 \cite{PhysRevLett.102.136806}. Subsequently, this physical phenomenon of a disorder-induced topological phase has been extensively studied,
including, but not limited to, Chern insulators \cite{PhysRevLett.105.115501,Zhang_2013CPB,PhysRevB.92.085410,PhysRevLett.116.066401,PhysRevB.100.054108},
topological insulators \cite{PhysRevB.80.165316,PhysRevLett.103.196805,PhysRevLett.105.216601,
PhysRevB.84.035110,PhysRevLett.113.046802,PhysRevB.91.214202,PhysRevB.96.205304,Wu_2016CPB},
topological superconductors \cite{PhysRevB.93.125133,Qin2016SR,PhysRevB.103.115430,Habibi2018PRB,Lieu2018PRB,PhysRevB.100.205302},
and higher-order topological insulators \cite{PhysRevLett.125.166801,PhysRevB.103.085408,PhysRevLett.126.146802}. For instance, in crystalline systems, disorder-induced chiral MEMs in 2D TSCs \cite{PhysRevB.93.125133,Qin2016SR,PhysRevB.103.115430} and disorder-induced MEMs in one-dimensional (1D) Kitaev superconductor chains \cite{Habibi2018PRB,Lieu2018PRB,PhysRevB.100.205302} have been reported in previous works. Meanwhile, it is important to note that the topological Anderson insulators can also be implemented in quasicrystalline systems, such as the disorder-induced 2D quantum spin Hall insulators in the Penrose tiling \cite{PhysRevB.100.115311} and AB tiling \cite{PhysRevB.103.085307} QLs, in which the topologically nontrivial phase is characterized by the nonzero spin Bott index and the quantized two-terminal conductance. However, the disorder-induced topological phases in QLs have still not been revealed in other topological classification systems, such as the TSCs in classes D and DIII. In view of the recent research on TSCs in QLs and the significant progress on disorder-induced topological phases in various systems, an intriguing question is whether disorder-induced MEMs can also emerge in the 2D quasicrystalline TSCs.

In this work, we systematically investigate the effects of Anderson-type disorder on the 2D AB tiling quasicrystalline TSCs, covering a chiral TSC in class D and a TRI TSC in class DIII. The AB tiling quasicrystal \cite{grunbaum1987tilings,c641bbfabe714fefbd7c37e571cb29fa,Duneau_1989} is tiled using squares and rhombuses with a small angle $45^{\circ}$ (see Fig.~\ref{fig1}). The construction process for the AB tiling quasicrystal through the inflation method was shown in Ref.~\cite{PhysRevLett.116.257002}. The Bott index (a real-space $\mathbb{Z}$ index for class D systems) and the spin Bott index (a real-space $\mathbb{Z}_2$ index for class DIII systems) are involved in evincing the topologically nontrivial phases with the MEMs in the class D chiral and class DIII TRI TSC systems. Meanwhile, to verify the results of the topological invariants, we bring in the recursive Green's function method for calculating the thermal conductance to test the existence of the MEMs in the two quasicrystalline TSC systems. We reveal rich phase diagrams of the two TSC systems when the disorder is turned on, and we find that the chiral and helical MEMs are stable for weak disorder, while strong disorder takes the MEMs away. We also show that a disorder-induced topologically nontrivial phase at certain parameter values in the class D chiral TSC system appears, accompanied by the disorder-induced chiral MEMs located at the square edge of the finite QL sample. Similarly, the disorder-induced helical MEMs can also be found in the class DIII quasicrystalline TSC.

The rest of this paper is organized as follows. In Sec.~\ref{Model}, we introduce two lattice tight-binding models with disorder on the AB tiling QL. Then, we give the details of numerical methods in Sec.~\ref{Methods} and reveal the chiral and helical MEMs of the two TSC systems in Sec.~\ref{without_Disorder}. Subsequently, in Sec.~\ref{Disorder}, we provide numerical results to study the topological phase transitions of the two TSC systems with disorder, and we end with a summary in Sec.~\ref{Conclusion}.

\section{Models}
\label{Model}
\begin{figure}[t]
	\includegraphics[width=5cm]{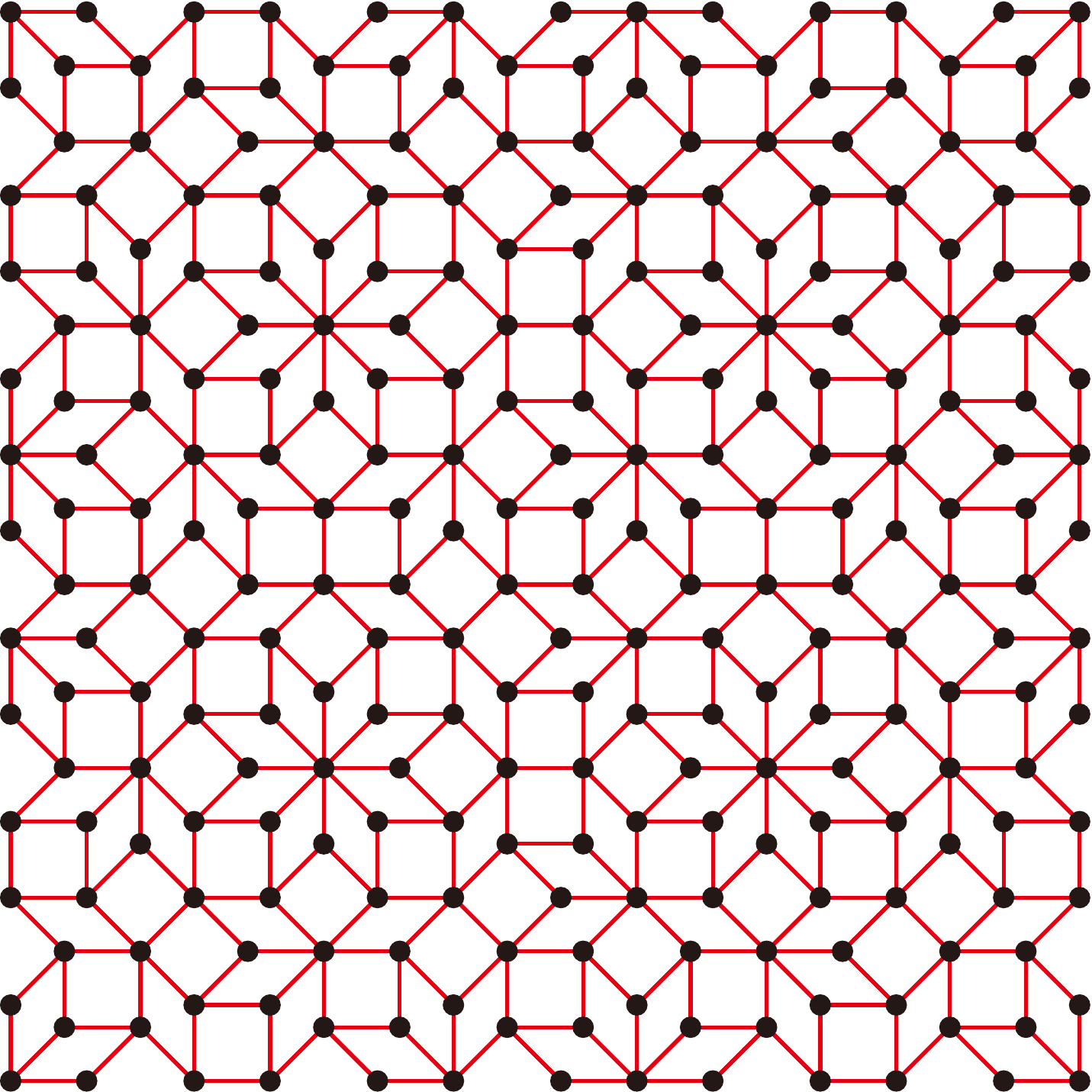} \caption{Schematic illustration of the Ammann-Beenker tiling quasicrystal. The quasicrystal consists of two types of primitive tiles: square tiles and rhombus tiles with a small angle  $45^{\circ}$. The black vertices represent the quasicrystal lattice sites. The lattice site connections of the short diagonal of the rhombus represent the nearest-neighbor bond, and the site connections of the sides of the two primitive tiles (red lines) represent the next-nearest-neighbor bond.}%
	\label{fig1}
\end{figure}

We start with two lattice tight-binding models which respectively describe a chiral TSC and a TRI TSC, with Anderson-type disorder on the AB tiling QL; a diagrammatic sketch of the QL is shown in Fig.~\ref{fig1}. Here, we discuss only the first two nearest-neighbor hopping and pairing terms (namely, only the nearest-neighbor and next-nearest-neighbor bonds in the QL are considered) and ignore the other long-range terms. In addition, we assume that the lattice site number is $N$, and the lattice site distance of the next-nearest-neighbor bond is used as the length unit. The class D Hamiltonian of the chiral TSC is given by
\begin{align}
H_{\text{D}}=& \sum_{j}\mu_{j} \psi_{j}^{\dag } \tau _{z}\psi_{j}+\sum_{j\not=k}\frac{u (d_{jk})}{2}\psi_{j}^{\dag } \left\{-t\tau _{z} \right. \nonumber \\
& \left.+i\Delta [\cos(\theta_{jk}) \tau _{x}+\sin( \theta_{jk})\tau _{y}] \right\}\psi_{k},
\label{HD}
\end{align}
where the basis is $\psi_{j}^{\dag }=(\varphi_{j}^{\dag }, \varphi_{j})$, and $j$ and $k$ denote lattice sites running from $1$ to $N$. $\tau_{x,y,z}$ are the Pauli matrices acting on the particle-hole degree of freedom. $t$ is the hopping amplitude, and $\Delta$ is the strength of the $p$-wave superconducting pairing. $\theta_{jk}$ is the polar angle of bond connecting sites $j$ and $k$ with respect to the horizontal direction. $u(d_{jk})=e^{-(d_{jk}-d_0)/\xi }$ is the spatial decay factor of the hopping and pairing terms, where $\xi$ is the decay length, $d_{jk}$ is the lattice site distance, and $d_0$ is the lattice site distance of the next-nearest-neighbor bond. The lattice site distance is $d_{jk}=\left\vert \mathbf{d}_{j}-\mathbf{d}_{k}\right\vert$, where $\mathbf{d}_{j}$ and $\mathbf{d}_{k}$ are the coordinates of the lattice sites. The Anderson-type disorder term is $\mu _{j}=\mu +W\omega _{j}$, where $\mu$ is the chemical potential, $\omega_{j}$ is the uniform random variable chosen from $\left[ -0.5,0.5\right]$, and $W$ is the disorder strength. The class D Hamiltonian (\ref{HD}) obeys only the PHS and satisfies the relation $P_{1}H_{\text{D}}P_{1}^{-1}=-H_{\text{D}}$. $P_{1}=\tau _{x}\mathcal{I}K$ is the PHS operator, where $K$ is the complex conjugate operator and $\mathcal{I}$ is the $N\times N$ identity matrix.

The class DIII Hamiltonian of the TRI TSC is written as
\begin{align}
H_{\text{DIII}}=& \sum_{j}\mu_{j} \psi_{j}^{\dag } \tau _{z}\sigma _{0}\psi_{j}+\sum_{j\not=k}\frac{u (r_{jk})}{2}\psi_{j}^{\dag } \left\{-t\tau _{z} \sigma _{0} \right. \nonumber \\ \phantom{=;\;\;}
& \left.+i\Delta \left[\cos(\theta_{jk}) \tau _{x} \sigma _{z}+\sin( \theta_{jk})\tau _{y}\sigma _{0} \right] \right\}\psi_{k},
\label{HDIII}
\end{align}
where the basis is $\psi_{j}^{\dag }=(\varphi_{j,\uparrow}^{\dag }, \varphi_{j,\uparrow}, \varphi_{j,\downarrow}^{\dag }, \varphi_{j,\downarrow})$. $\sigma_{0}$ and $\tau_{0}$ are the $2\times2$ identity matrices. $\sigma_{x,y,z}$ and $\tau_{x,y,z}$ are the Pauli matrices acting on the spin and particle-hole degrees of freedom, respectively. Other physical quantities have the same physical meaning as for the class D Hamiltonian $H_{\text{D}}$. The class DIII Hamiltonian (\ref{HDIII}) satisfies
\begin{align}
P_{2}H_{\text{DIII}}P_{2}^{-1}&=-H_{\text{DIII}},  \nonumber \\
TH_{\text{DIII}}T^{-1}&=H_{\text{DIII}},  \\
CH_{\text{DIII}}C^{-1}&=-H_{\text{DIII}}. \nonumber
\end{align}
Here, $P_{2}$, $T$, and $C$ are the PHS, TRS, and chiral symmetry operators, respectively, and they are expressed by
\begin{equation}
P_{2}=\tau _{x}\sigma _{0}\mathcal{I}K, T=\tau _{0}\sigma _{y}\mathcal{I}K, C=P_{2} T.
\end{equation}
Herein, the energy units are regulated as $t$, and the spatial decay length $\xi$ and the lattice site distance $d_0$ are set as $1$.

\section{Numerical Methods}
\label{Methods}

\subsection{Bott index and spin Bott index}

The topologically nontrivial phase with edge states is characterized by the bulk topological invariant. Here, we briefly introduce the two real-space topological invariants, which are employed in characterizing the topological phases of the two TSCs in the AB tiling QL because the QL lacks translational symmetry and cannot be handled by the momentum-space topological invariants. The two real-space topological invariants are the Bott index \cite{Loring_2010EPL,HASTINGS20111699,LORING2015383,PhysRevX.6.011016,PhysRevB.98.125130,ghadimi2020topological}, a $\mathbb{Z}$ topological index to characterize the 2D chiral TSC system with chiral MEMs in class D, and the spin Bott index \cite{PhysRevLett.121.126401,PhysRevB.98.125130,PhysRevB.100.115311,PhysRevB.103.085307,PhysRevResearch.3.033227}, a $\mathbb{Z}_2$ topological index to characterize the 2D TRI TSC system with helical MEMs in class DIII. It is noted that the Bott index and the spin Bott index are both numerically calculated in real space with the approximate periodic boundary condition, in which a square-shaped AB tiling QL is transformed to a torus geometry.

First of all, we detail the concrete numerical calculation steps for the Bott index \cite{Loring_2010EPL,HASTINGS20111699,LORING2015383,PhysRevX.6.011016,PhysRevB.98.125130,ghadimi2020topological}. Initially, we establish the occupation projector operator as
\begin{align}
 Q=\sum_{j}^{L}|\Psi _{j}\rangle \langle \Psi _{j}|,
\label{P}
\end{align}
where $\Psi _{j}$ is the $j$th eigenvector of the Hamiltonian and $j$ runs from $1$ to $L$. $L$ is the total number of negative eigenvalues, where the negative energy states are the occupied states owing to the PHS of the TSC systems. Then, we define the projected position operators as
\begin{align}
&U_{X}=Q e^{i2\pi X}Q+(I-Q),\\
&V_{Y}=Qe^{i2\pi Y}Q+(I-Q),
\end{align}
where $I$ is the $L\times L$ identity matrix. $X$ and $Y$ are two diagonal matrices, and the diagonal elements are $X_{jj}=x_{j}$ and $Y_{jj}=y_{j}$ with the coordinate  $(x_{j},y_{j})$ of the $j$th lattice site, where the coordinates are rescaled to the interval $[0,1)$. By means of gauging the commutativity of the projected position operators \cite{PhysRevB.98.125130}, the Bott index is defined as
\begin{align}
B=\frac{1}{2\pi }{\rm Im}\{{\rm Tr}[\ln (V_{Y}U_{X}V_{Y}^{\dag}U_{X}^{\dag })]\}.
\label{Bott}
\end{align}
The case with $B=0$ corresponds to the topologically trivial phase with no MEMs, and $B=1$ corresponds to the topologically nontrivial phase with chiral MEMs.

After explaining the numerical calculation method for the Bott index, we next construct the concrete numerical calculation steps for the spin Bott index \cite{PhysRevLett.121.126401,PhysRevB.98.125130,PhysRevB.100.115311,PhysRevB.103.085307}. First, we formulate the projected spin operator as
\begin{equation}
Q_{z}=Q\hat{\eta}_{z}Q,
 \end{equation}
 where $Q$ is the occupation projector operator and $\hat{\eta}_{z}=\frac{\hbar }{2}\sigma _{z}$ is the spin operator with the Pauli matrix $\sigma _{z}$. The eigenvalues of $Q_z$ still remain two isolated parts divided by zero energy. Then, we define a new projector operator as
\begin{align}
Q_{\pm }=\sum_{j}^{L/2}\left\vert \Phi _{j}^{\pm }\right\rangle \left\langle \Phi _{j}^{\pm }\right\vert,
\label{P1}
\end{align}
where $\Phi _{j}^{+}$ ($\Phi _{j}^{-}$) is the eigenvector corresponding to the $j$th positive (negative) eigenvalue of $Q_z$. Subsequently, we formulate
the new projected position operators as
\begin{align}
&U_{\pm }=Q_{\pm }e^{i2\pi X}Q_{\pm }+(I-Q_{\pm }),\\
&V_{\pm }=Q_{\pm }e^{i2\pi Y}Q_{\pm }+(I-Q_{\pm }).
\label{UV}
\end{align}

In numerical calculations, we adopt the singular value decomposition method to calculate the projected position operators $U_{\pm }$ and $V_{\pm }$ to improve the stability of the numerical results of the spin Bott index. The singular value decomposition of a matrix can be expressed as $M=Z\Lambda \Pi ^{\dag }$, where $Z$ and $\Pi$ are unitary and $\Lambda$ is real and diagonal. We specify the ``unitary part" $\tilde{M}=Z\Pi ^{\dag }$ as the new projected position operator and replace the initial matrix $M$ \cite{PhysRevB.98.125130}. Finally, the spin Bott index can be defined as
\begin{align}
B_{s}=\frac{1}{2}(B_{+}-B_{-}),
\label{Bott}
\end{align}
with
\begin{align}
B_{\pm }=\frac{1}{2\pi }{\rm Im}\{{\rm Tr}[\ln (\tilde{V}_{\pm }\tilde{U}_{\pm }\tilde{V}_{\pm }^{\dag
}\tilde{U}_{\pm }^{\dag })]\},
\label{Bott}
\end{align}
where $B_{\pm }$ are the Bott indexes of two spin sectors. The case with $B_{s}=0$ corresponds to the topologically trivial phase with no MEMs, and $B_{s}=1$ corresponds to the topologically nontrivial phase with helical MEMs.

\subsection{Thermal conductance}

Meanwhile, we test the topological nature of the MEMs of the chiral and TRI TSCs in the AB tiling QL by studying the thermal transport properties of the systems. The setup, a two-terminal normal-metal--superconductor--normal-metal (NSN) junction, is constructed by attaching two semi-infinite normal metal leads to the left and right ends of the superconductor device. The normal-metal lead Hamiltonian is described by the superconductor device Hamiltonian in a square lattice, where $W$ and $\Delta$ are set to zero. The connected Hamiltonian, which represents the connection of the normal-metal lead and the superconductor device, is described by the superconductor device Hamiltonian by setting $\mu$, $W$, and $\Delta$ to zero. Here, the hopping amplitudes of the superconductor device Hamiltonian, the normal-metal lead Hamiltonian, and the connected Hamiltonian are all set to be equal.

In order to compute the two-terminal thermal conductance, we first calculate the Fermi level ($E=0$) scattering matrix $S$ of the NSN junction with
\begin{align}
S=\left(
\begin{array}{cc}
S_{\rm LL} & S_{\rm LR} \\
S_{\rm RL} & S_{\rm RR}
\end{array}
\right),
\end{align}
where the block matrices $S_{\rm LL}$ and $S_{\rm RR}$ represent the reflection amplitudes, the block matrices $S_{\rm LR}$ and $S_{\rm RL}$ represent the transmission amplitudes, and the subscripts L and R represent left and right leads, respectively. Each block of the scattering matrix is
\begin{align}
S_{mn}=\left(
\begin{array}{cc}
S_{mn}^{ee} & S_{mn}^{e\rm h} \\
S_{mn}^{{\rm h} e} & S_{mn}^{\rm hh}%
\end{array}
\right),
\end{align}
where $m,n=\rm L, \rm R$. The element $S_{mn}^{\alpha\beta}$ indicates the scattering amplitude of an outgoing $\beta $ particle attributed to the incoming $\alpha $ particle, where $\alpha$ and $\beta$ denote the electron ($e$) or hole (h). The scattering matrix is numerically calculated by utilizing the recursive Green's function method \cite{MacKinnon_1985,PhysRevB.72.235304,PhysRevB.86.174512,Lewenkopf2013JCE,
PhysRevB.88.064509,PhysRevB.100.205302}. The scattering matrix $S$, related to the Green's function, is given by \cite{Lee1981PRL,Fisher1981PRB}
\begin{equation}
S_{mn}^{\alpha\beta}=-\delta_{m,n}\delta_{\alpha,\beta}+i\left[\Gamma_{m}^{\alpha}\right]^{1/2}G^{r}\left[\Gamma_{n}^{\beta }\right]^{1/2}.
\end{equation}
$\Gamma_{m}^{\alpha }$ is the linewidth function of the $\alpha $ particle with $\Gamma_{m}^{\alpha }=i[(\Sigma_{m}^{\alpha})^{r}-(\Sigma_{m}^{\alpha})^{a}]$, where $(\Sigma_{m}^{\alpha})^{r/a}$ is the retarded (advanced) self-energy of the $\alpha $ particle for the $m$ lead. $G^{r}$ is the retarded Green's function of the superconductor device and can be expressed as
\begin{equation}
G^{r}=\left[E+i0^{+}-H-\sum_{m,\alpha} ( \Sigma_{m}^{\alpha})^{r}\right] ^{-1},
\end{equation}
where $H$ is the superconductor device Hamiltonian and $E$ is the Fermi level and is set to zero.

Therefore, in the low-temperature linear response regime, the thermal conductance is formulated as \cite{RevModPhys.87.1037}
\begin{equation}
G=G_{0} \rm Tr(S_{\rm LR}^{\dagger} S_{\rm LR}),
\end{equation}
with the quantum of thermal conductance $G_{0}=\pi^2 k_{B}^{2} T_0 /6h$. The quantized thermal conductance $G/G_{0}=1$ is the signature of a chiral MEM located at the edge of the superconductor device, and $G/G_{0}=0$ if there is no MEM. The thermal conductance $G/G_{0}=2$ indicates that there is a pair of MEMs, the helical MEMs, located at the edge of the superconductor device.

\section{chiral and helical MEMs in clean limit}
\label{without_Disorder}
\begin{figure}[t]
	\includegraphics[width=8.5cm]{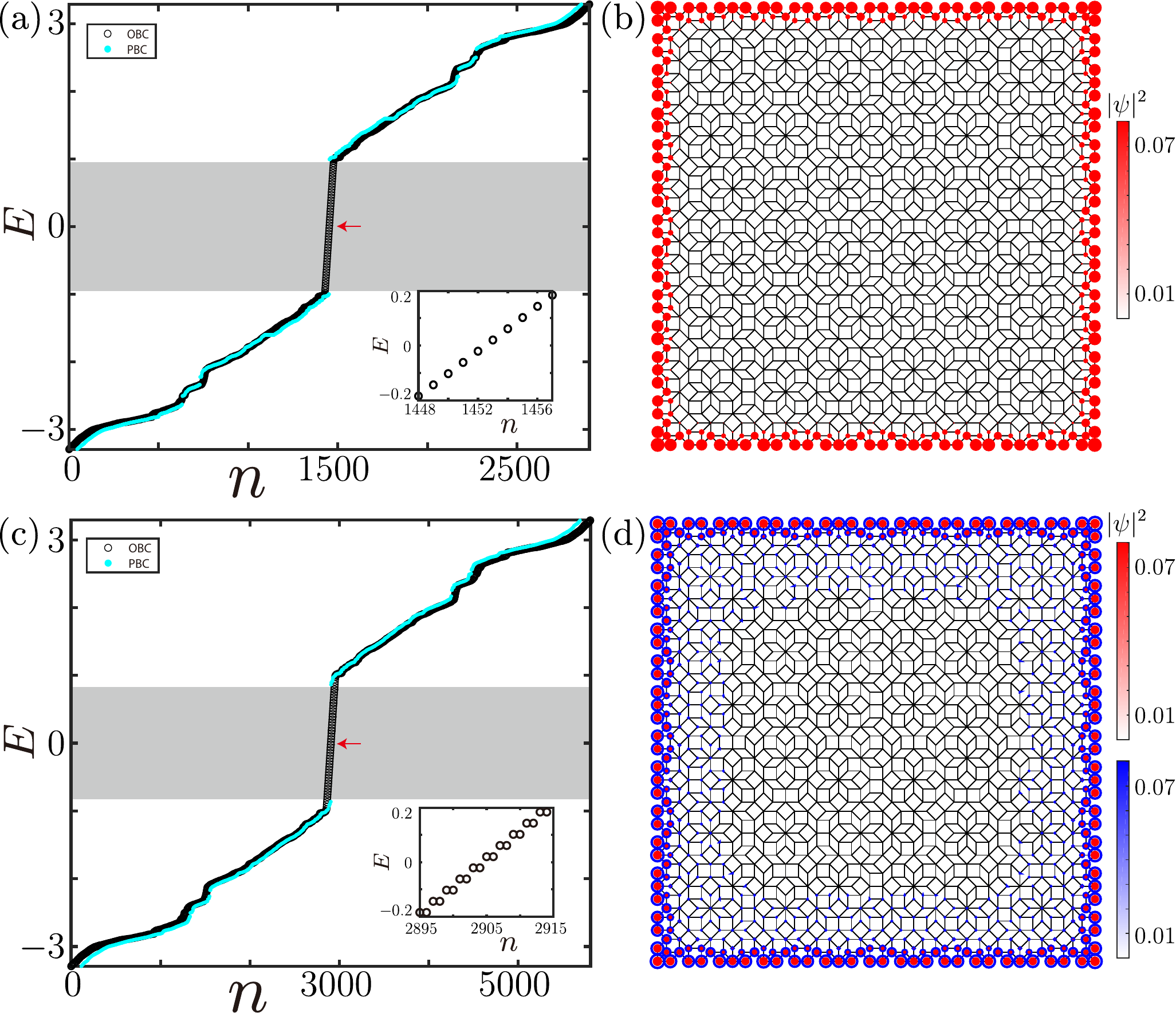} \caption{(a) Energy spectrum of the class D Hamiltonian $H_{\text{D}}$ on the AB tiling QL with a square shape under the OBC (black circles) and the PBC (cyan dots) versus the eigenvalue index $n$. The inset shows the enlarged section of eigenstates near zero energy for the system under the OBC. The gray region shows the midgap states. (b) The probability density of the in-gap eigenstates near zero energy is marked by the red arrow in (a). The color map shows the values of the probability density. (c) Energy spectrum of the class DIII Hamiltonian $H_{\text{DIII}}$ on the AB tiling QL with a square shape under the OBC (black circles) and the PBC (cyan dots) versus the eigenvalue index $n$. (d) The probability density of doubly degenerate in-gap eigenstates near zero energy is marked by the red arrow in (c). The red dots (blue circles) represents an edge state with spin up (down). We take the model parameters $\Delta/t=1$, $\mu/t=1.6$, $W/t=0$, and lattice site number $N=1452$.}%
	\label{fig2}
\end{figure}

In this section, to reveal the topologically nontrivial phase with MEMs in the clean limit, we directly diagonalize the class D Hamiltonian~(\ref{HD}) and the class DIII Hamiltonian~(\ref{HDIII}) on the AB tiling QL with square geometry under an open boundary condition (OBC) and a periodic boundary condition (PBC), respectively. Here, we set the model parameters $\Delta/t=1$, $\mu/t=1.6$, $W/t=0$, and lattice site number $N=1452$.

Figure~\ref{fig2}(a) shows the energy spectrum of the class D Hamiltonian~(\ref{HD}) under the OBC (black circles) and the PBC (cyan dots) versus the eigenvalue index $n$. It is found that an energy gap for the PBC system emerges in the energy spectrum, while the gapless in-gap energy states for the OBC system fill the bulk energy gap of the PBC system. In Fig.~\ref{fig2}(b), we plot the probability density of an in-gap eigenstate near zero energy [marked by the red arrow in Fig.~\ref{fig2}(a)] for a finite QL sample with square boundary geometry under the OBC. Interestingly enough, we find that the in-gap state is located at the square edge of the finite QL sample. We further calculate the Bott index to identify the topological origin of the edge modes. In the case of the same parameters as in Fig.~\ref{fig2}(a), the numerical calculation result of the Bott index is $B=1$, which indicates that this phase is topologically nontrivial with chiral MEMs. Meanwhile, the two-terminal thermal conductance, where $G/G_{0}=1$, is calculated to confirm the topological nature of the edge modes.

Similarly, we show the energy spectrum of the class DIII Hamiltonian~(\ref{HDIII}) under the OBC (black circles) and the PBC (cyan dots) versus the eigenvalue index $n$ in Fig.~\ref{fig2}(c). We also find that the PBC system possesses an energy gap, while the OBC system has gapless in-gap states occupying the bulk energy gap of the PBC system. Note that all the in-gap states are doubly degenerate states due to the TRS. Figure~\ref{fig2}(d) shows the probability density of an in-gap eigenstate near zero energy [marked by the red arrow in Fig.~\ref{fig2}(c)] for a finite QL sample with square boundary geometry under the OBC. The red dots (blue circles) represent an edge state with spin up (down). It is found that the in-gap states are located at the square edge of the finite QL sample. The topologically nontrivial phase of a TRI TSC with the helical MEMs is confirmed by the numerical results of the spin Bott index $B_{s}=1$ and the two-terminal thermal conductance $G/G_{0}=2$.

\section{the effects of disorder}
\label{Disorder}

In this section, we numerically investigate the effects of the Anderson-type disorder on the topological phase transitions of the chiral and TRI quasicrystalline TSC systems. Based on the calculation of the real-space topological invariants and the two-terminal thermal conductance, topological phase diagrams with different parameters will be presented.

\subsection{Class D}

\begin{figure}[t]
	\includegraphics[width=8.5cm]{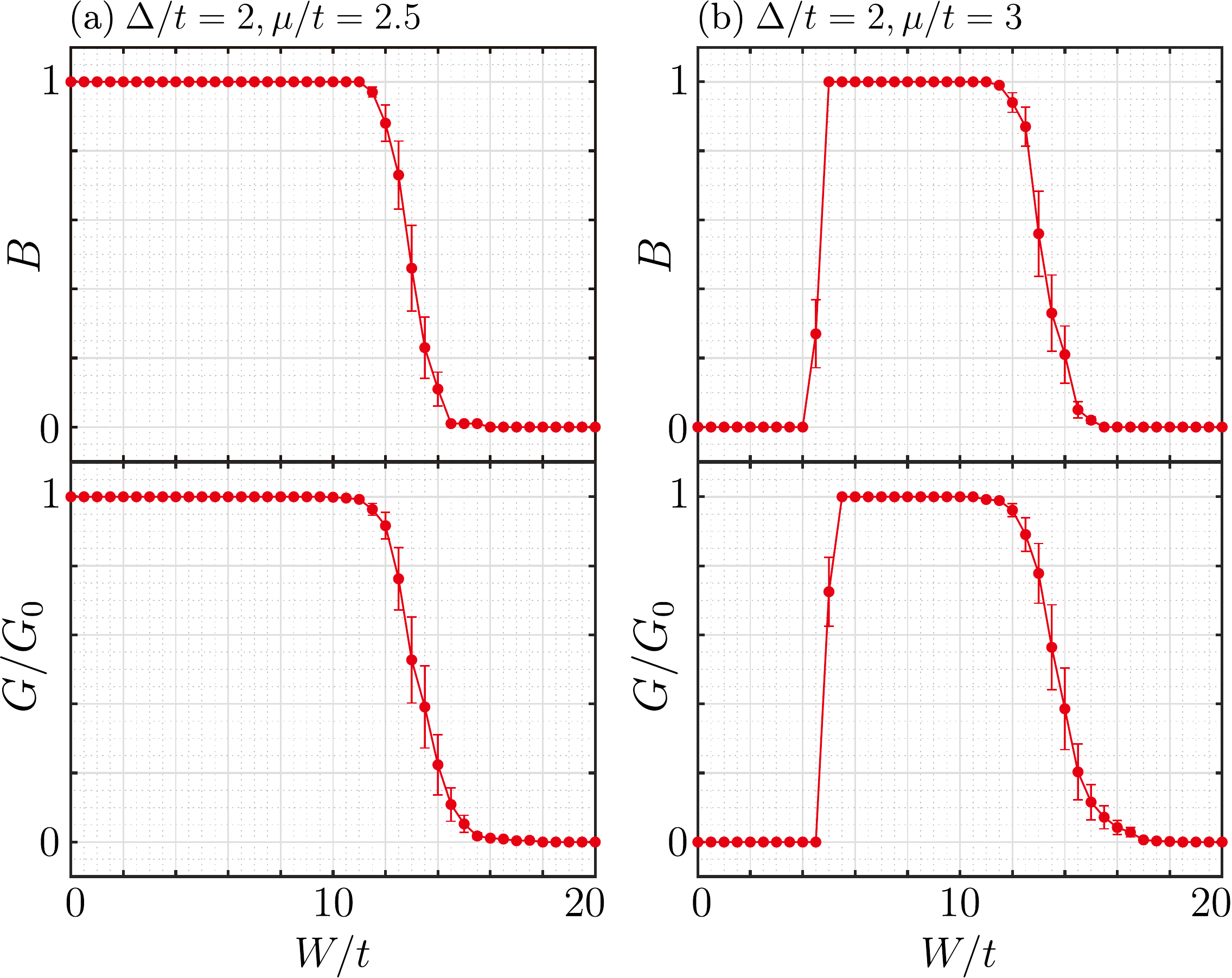} \caption{The Bott index $B$ and the thermal conductance $G/G_{0}$ for the class D TSC system as a function of the disorder strength $W/t$ for (a) $\mu/t=2.5$ and (b) $\mu/t=3$. We take the parameter $\Delta/t=2$. In calculating the Bott index (the thermal conductance), the lattice site number of the QL is taken to be $N=1452$ ($8260$), and the error bar indicates a standard deviation of $500$ ($1000$) samples.}%
	\label{fig3}
\end{figure}

First, we reveal the disorder-induced topological phase transitions in the class D chiral TSC system. First of all, based on the computation of the Bott index $B$ and the two-terminal thermal conductance $G/G_0$, we study the effects of disorder on the topological phase transitions for two sets of system parameters. For the case of $(\Delta/t,\mu/t)=(2,2.5)$, the phase is topologically nontrivial with nonzero Bott index $B=1$ in the clean limit. Figure~\ref{fig3}(a) shows the Bott index $B$ and the thermal conductance $G/G_0$ in this case as a function of the disorder strength $W/t$. We find that the topologically nontrivial phase remains stable when the disorder strength is small, which is characterized by the nonzero Bott index $B=1$ and the quantized thermal conductance $G/G_{0}=1$ in a certain range of disorder strength ($0\leq W/t\leq11$). However, with the disorder strength $W/t$ increasing, a topological phase transition occurs at $W/t=11$, beyond which both the Bott index $B$ and the thermal conductance $G/G_0$ decay to zero, and the class D chiral TSC system is converted to a topologically trivial phase.

For the case of $(\Delta/t,\mu/t)=(2,3)$, the phase is topologically trivial with zero Bott index, $B=0$, in the clean limit. The Bott index $B$ and the thermal conductance $G/G_0$ in this case as a function of the disorder strength ($W/t$) are plotted in Fig.~\ref{fig3}(b). With the increase in $W/t$, it is found that two topological phase transitions arise, accompanied by the Bott index changing from $B=0$ to $B=1$ at $W/t=5$ and returning to $B=0$ at $W/t=11$. Here, plateau of the nonzero Bott index, $B=1$, exists in a certain range of disorder strength ($5\leq W/t\leq11$), which indicates a topologically nontrivial phase induced by disorder. Meanwhile, the numerical result of the thermal conductance is obtained, and we find that it can match well the numerical result of the Bott index. The value of the thermal conductance jumps from $G/G_0=0$ to $G/G_0=1$ at $W/t=5.5$ and goes back to $G/G_0=0$ at $W/t=11$. Thus, the chiral MEMs can be induced by disorder when the disorder strength is in the region of $5.5\leq W/t\leq11$ in the class D chiral TSC system (with model parameters $\Delta/t=2$, and $\mu/t=3$).

\begin{figure}[t]
	\includegraphics[width=8.5cm]{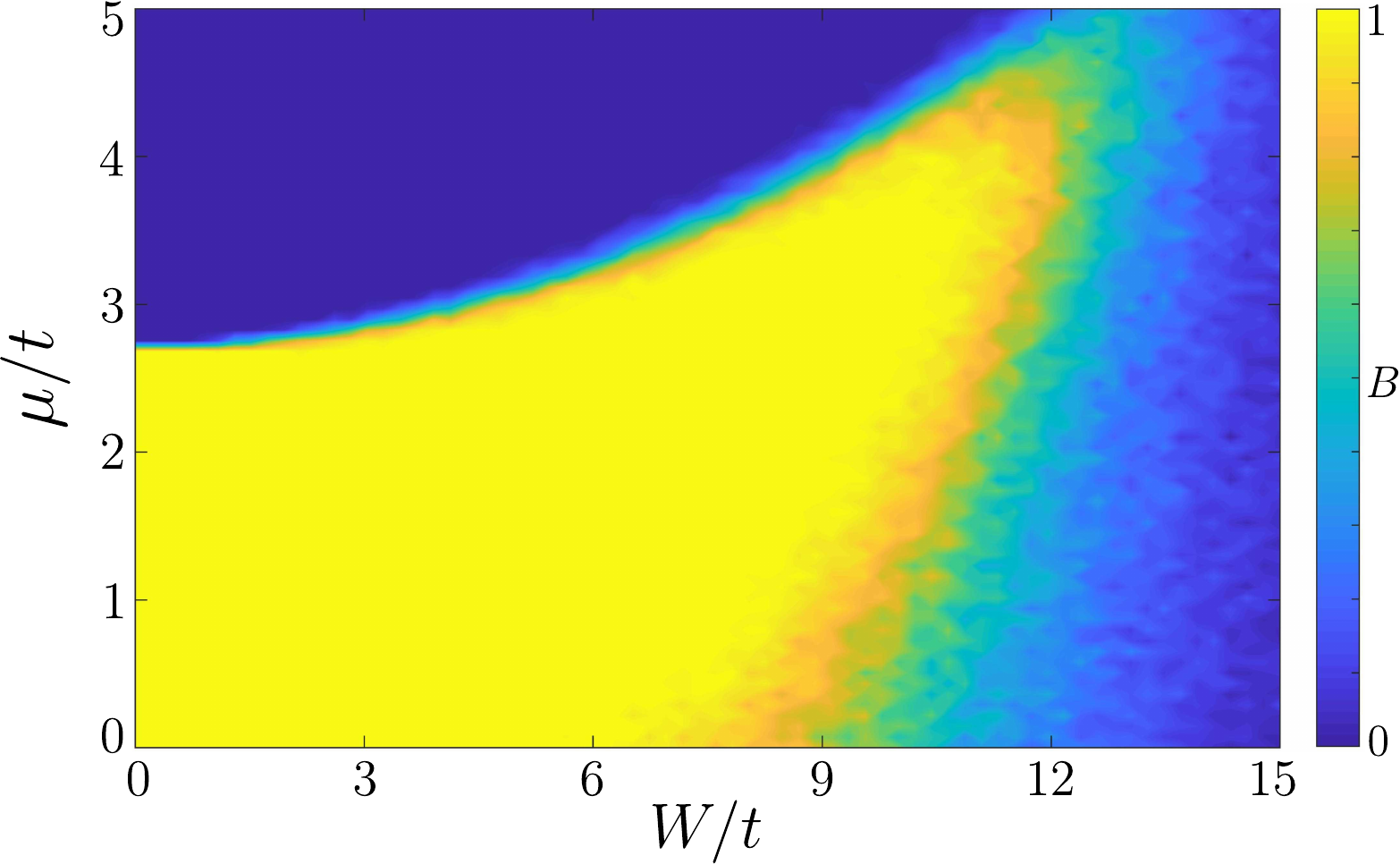} \caption{Phase diagram in ($W/t$, $\mu/t$) space for the class D TSC system with disorder obtained by calculating the Bott index $B$ with $100$ disorder configurations. The yellow region denotes the topologically nontrivial phase ($B=1$), and the blue region denotes the topologically trivial phase ($B=0$). We take the parameters $\Delta/t=2$ and $N=264$.}%
	\label{fig4}
\end{figure}

Additionally, the topological phase diagram for the class D system with disorder in the ($W/t$, $\mu/t$) space is plotted in Fig.~\ref{fig4}, where $\Delta/t=2$. The color map shows the values of the Bott index $B$. The yellow region denotes the topologically nontrivial phase with $B=1$, and the blue region denotes the topologically trivial phase with $B=0$. It is found that the maximum disorder strength, below which the topologically nontrivial phase remains stable, increases with increasing the chemical potential $\mu/t$. The largest maximum disorder strength is about $W/t\approx12$, beyond which the topologically nontrivial phase vanishes. We also find the disorder-induced topologically nontrivial phase region in a range of parameters ($W/t$, $\mu/t$) space is distinctly presented in the topological phase diagram, as shown in Fig.~\ref{fig4}.

\begin{figure}[t]
	\includegraphics[width=8.5cm]{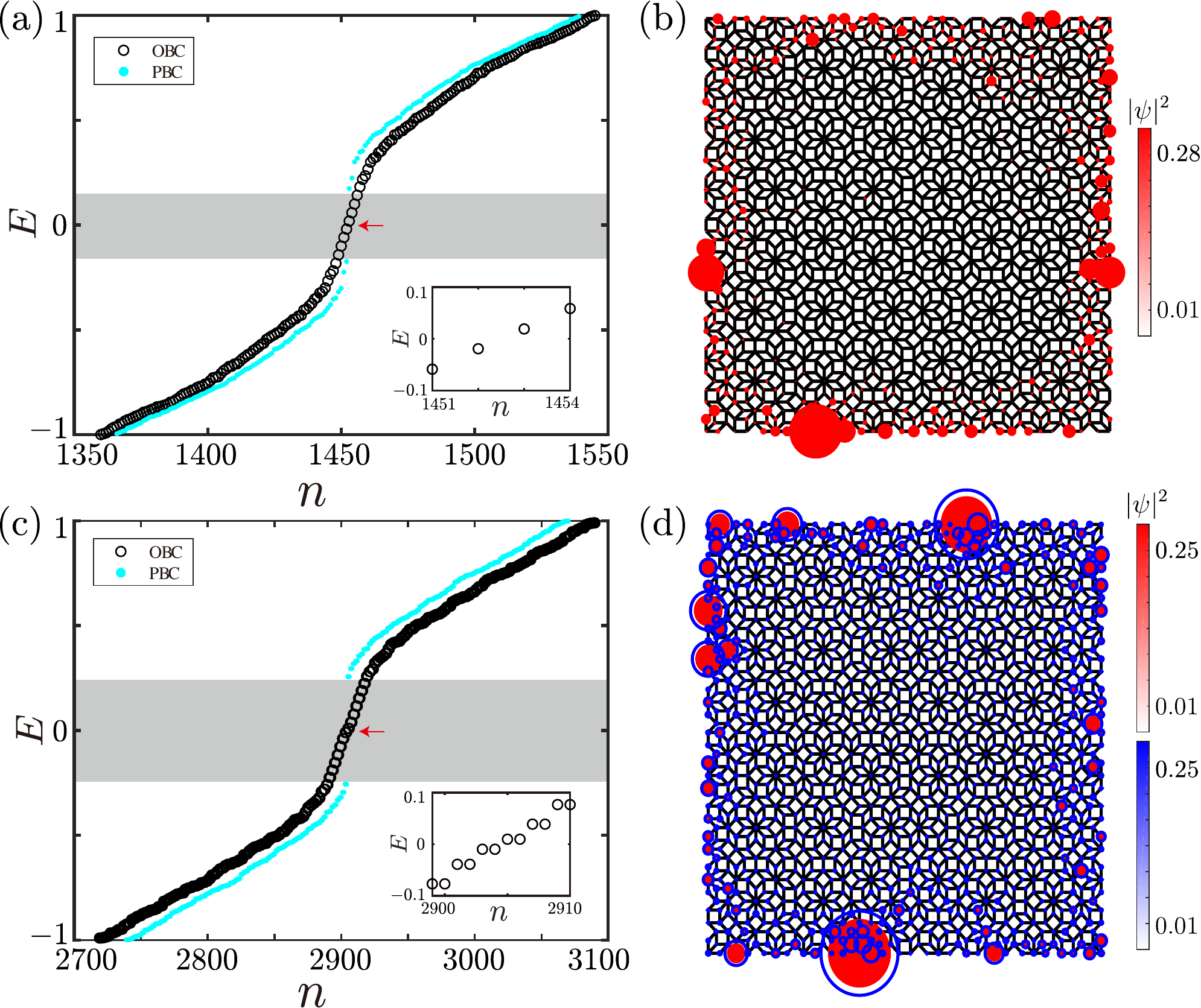} \caption{(a) Energy spectrum of the class D Hamiltonian $H_{\text{D}}$ on the AB tiling QL with a square shape under the OBC (black circles) and the PBC (cyan dots) versus the eigenvalue index $n$. The inset shows the enlarged section of eigenstates near zero energy for the system under the OBC. The gray region shows the midgap states. (b) The probability density of the in-gap eigenstates near zero energy is marked by the red arrow in (a). The color map shows the values of the probability density. (c) Energy spectrum of the class DIII Hamiltonian $H_{\text{DIII}}$ on the AB tiling QL with a square shape under the OBC (black circles) and the PBC (cyan dots) versus the eigenvalue index $n$. (d) The probability density of doubly degenerate in-gap eigenstates near zero energy is marked by the red arrow in (c). The red dots (blue circles) represent an edge state with spin up (down). We take the model parameters $\Delta/t=2$, $\mu/t=3$, $W/t=8$, and lattice site number $N=1452$.}%
	\label{fig5}
\end{figure}

In order to show the disorder-induced MEMs in the class D TSC system clearly, we directly diagonalize the class D Hamiltonian~(\ref{HD}) on the QL with square geometry under the OBC and PBC. Here, we set the model parameters $\Delta/t=2$, $\mu/t=3$, $W/t=8$, and lattice site number $N=1452$. Figure~\ref{fig5}(a) shows the energy spectrum of the class D Hamiltonian~(\ref{HD}) under the OBC (black circles) and the PBC (cyan dots) versus the eigenvalue index $n$. It is found that an energy gap for the PBC system emerges in the energy spectrum, while the gapless in-gap energy states for the OBC system fill the bulk energy gap of the PBC system. In Fig.~\ref{fig5}(b), we plot the probability density of an in-gap eigenstate near zero energy [marked by the red arrow in Fig.~\ref{fig5}(a)] for a finite QL sample with square boundary geometry under the OBC. Interestingly enough, we find that the in-gap state is located at the square edge of the finite QL sample.

\subsection{Class DIII}

\begin{figure}[t]
	\includegraphics[width=8.5cm]{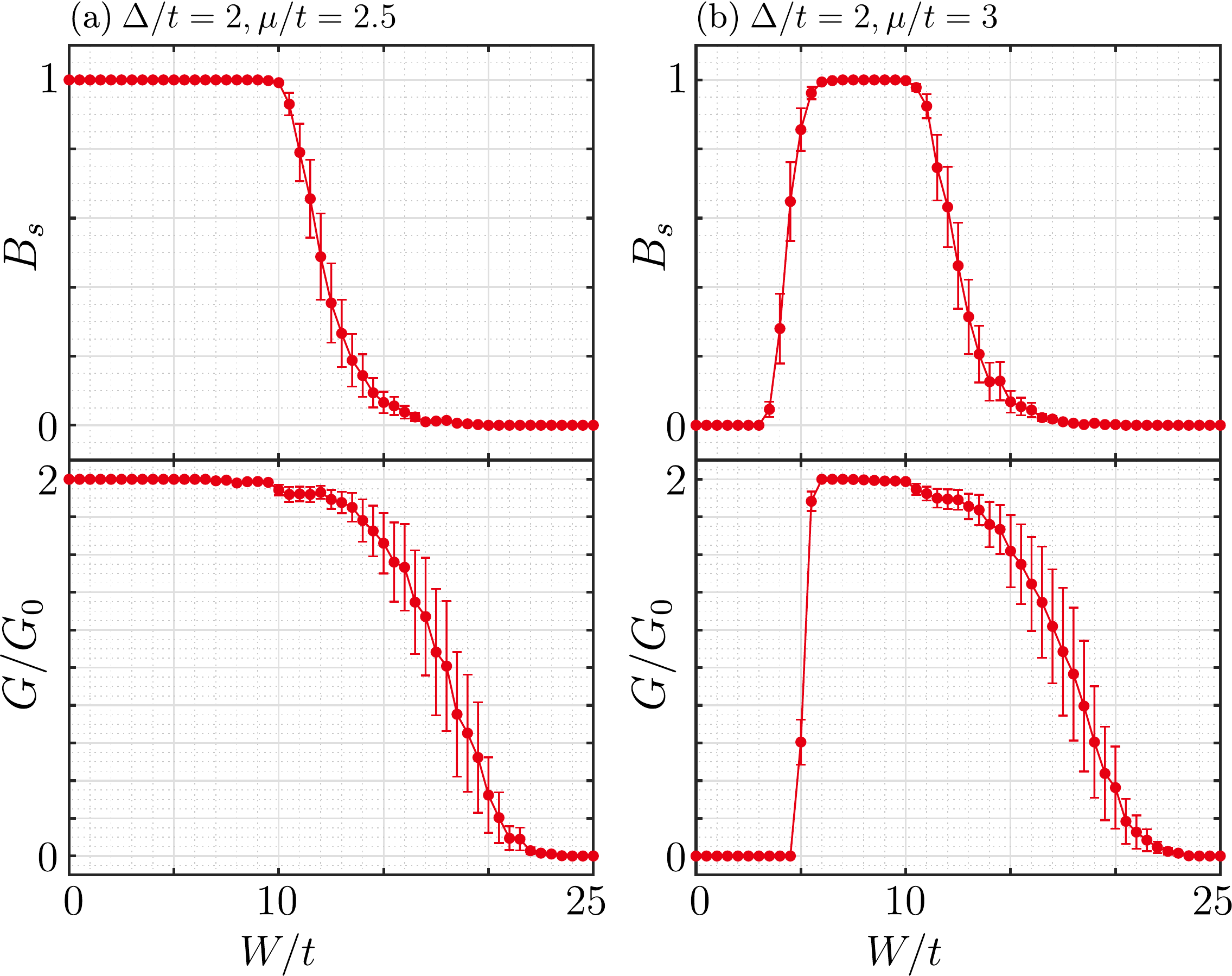} \caption{The spin Bott index $B_s$ and the thermal conductance $G/G_{0}$ for the class DIII TSC system as a function of the disorder strength $W/t$ for (a) $\mu/t=2.5$ and (b) $\mu/t=3$. We take the parameter $\Delta/t=2$. In calculating the spin Bott index (the thermal conductance), the lattice site number of the QL is taken to be $N=264$ ($8260$), and the error bar indicates a standard deviation of $500$ ($1000$) samples.}%
	\label{fig6}
\end{figure}

Next, we reveal the disorder-induced topological phase transitions in the class DIII TRI TSC system. Like for the class D case, we first study the effects of disorder on the topological phase transitions at two sets of system parameters based on the computation of the spin Bott index $B_s$ and the two-terminal thermal conductance $G/G_0$. Figure~\ref{fig6} shows the spin Bott index $B_s$ and the thermal conductance $G/G_0$ as a function of the disorder strength $W/t$.

For the case of $(\Delta/t,\mu/t)=(2,2.5)$, the phase is topologically nontrivial with the nonzero spin Bott index, $B_s=1$, in the clean limit. Figure~\ref{fig6}(a) shows that the topologically nontrivial phase remains stable when the disorder strength is small, which is characterized by the nonzero spin Bott index $B_s=1$ and the quantized thermal conductance $G/G_{0}=2$ in a certain range of disorder strength ($0\leq W/t\leq9.5$). Then, a topological phase transition occurs at $W/t=9.5$ with further increasing $W/t$, beyond which both the spin Bott index $B_s$ and the thermal conductance $G/G_0$ decay to zero, and the class DIII TRI TSC system is converted to a topologically trivial phase. For the case of $(\Delta/t,\mu/t)=(2,3)$, the phase is topologically trivial with zero spin Bott index, $B_s=0$, in the clean limit. Figure~\ref{fig6}(b) shows that two topological phase transitions arise with increasing $W/t$, accompanied by the spin Bott index changing from $B_s=0$ to $1$ at $W/t=6$ and returning to zero at $W/t=10$, and the thermal conductance jumping from $G/G_0=0$ to $2$ at $W/t=6$ and returning to zero at $W/t=10$. The plateaus of the nonzero spin Bott index $B_s=1$ and the quantized thermal conductance $G/G_0=2$ exist in a certain range of disorder strength ($6\leq W/t\leq10$), which indicates that a topologically nontrivial phase is induced by disorder. Thus, the helical MEMs can be induced by disorder when the disorder strength is in the region of $6\leq W/t\leq10$ in the class DIII TRI TSC system (with model parameters $\Delta/t=2$ and $\mu/t=3$).

\begin{figure}[t]
	\includegraphics[width=8.5cm]{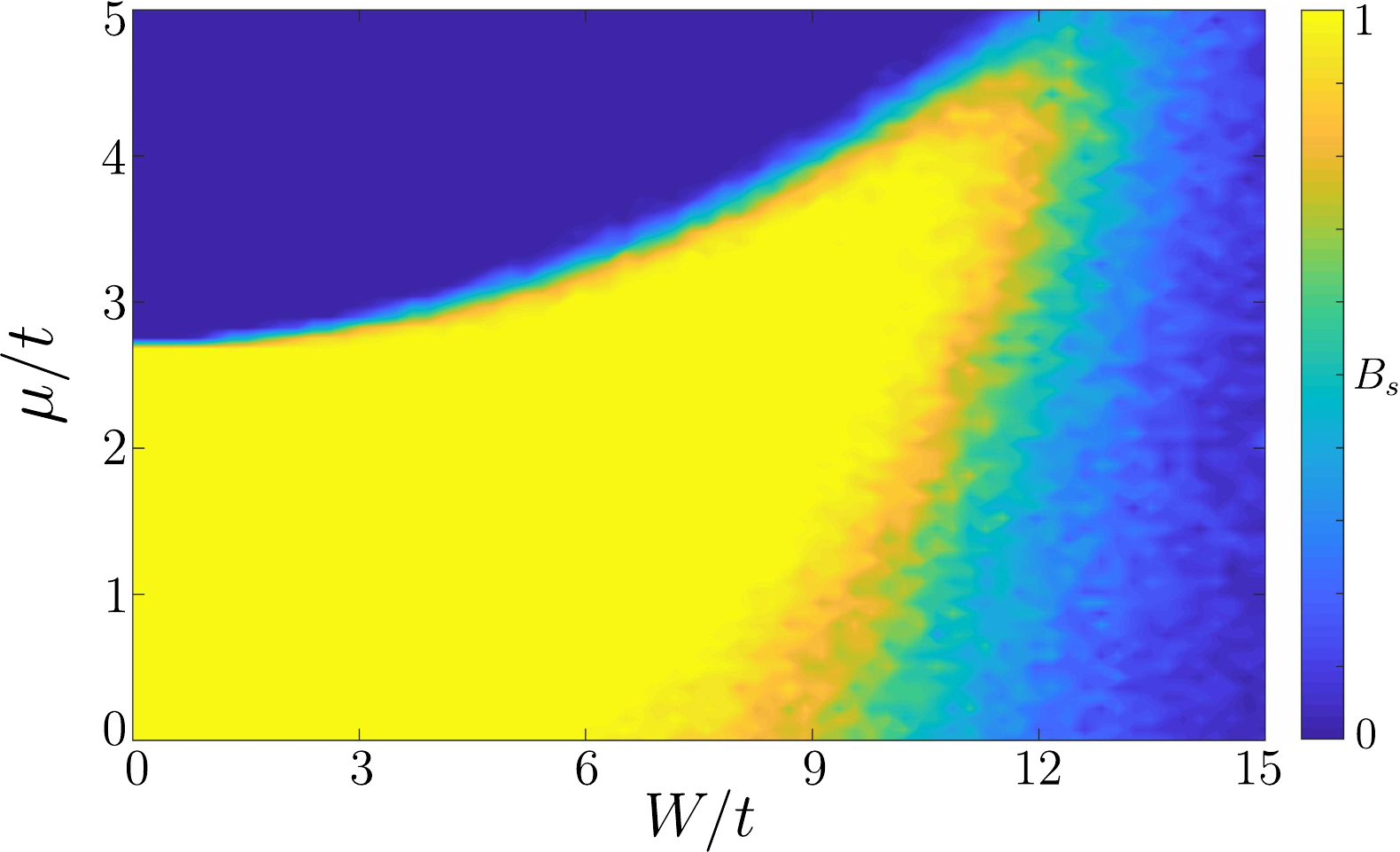} \caption{Phase diagram in ($W/t$, $\mu/t$) space for the class DIII TSC system with disorder obtained by calculating the spin Bott index $B_s$ with $100$ disorder configurations. The yellow region denotes the topologically nontrivial phase ($B_s=1$), and the blue region denotes the topologically trivial phase ($B_s=0$). We take the parameter $\Delta/t=2$ and $N=264$.}%
	\label{fig7}
\end{figure}

Additionally, the topological phase diagram for the class DIII TRI TSC system with disorder in the ($W/t$, $\mu/t$) space is plotted in Fig.~\ref{fig7}, where $\Delta/t=2$. The color map shows the values of the spin Bott index $B_s$. The yellow region denotes the topologically nontrivial phase with $B_s=1$, and the blue region denotes the topologically trivial phase with $B_s=0$. It is found that the largest maximum disorder strength is about $W/t\approx12$, beyond which the topologically nontrivial phase vanishes. We also find that the disorder-induced topologically nontrivial phase region, in a range of parameter ($W/t$, $\mu/t$) space, is distinctly presented in the topological phase diagram, as shown in Fig.~\ref{fig7}. It is noted that the disorder-averaged topological phase diagram (Fig.~\ref{fig7}) of class DIII is obviously identical to that (Fig.~\ref{fig4}) of class D. The reason is that the Hamiltonian~(\ref{HDIII}) of class DIII comprises the two-time-reversal partner Hamiltonian~(\ref{HD}) of class D, and moreover, here, the on-site disorders introduced do not break the time-reversal symmetry of the Hamiltonian~(\ref{HDIII}).

In order to show the disorder-induced MEMs in the class DIII TSC system clearly, we directly diagonalize the class DIII Hamiltonian~(\ref{HDIII}) on the QL with square geometry under the OBC and PBC. Here, we set the model parameters $\Delta/t=2$, $\mu/t=3$, $W/t=8$, and lattice site number $N=1452$. We show the energy spectrum of the class DIII Hamiltonian~(\ref{HDIII}) under the OBC (black circles) and the PBC (cyan dots) versus the eigenvalue index $n$ in Fig.~\ref{fig5}(c). We also find that the PBC system possesses an energy gap, while the OBC system has gapless in-gap states occupying the bulk energy gap of the PBC system. Note that all the in-gap states are doubly degenerate states due to the TRS. Figure~\ref{fig5}(d) shows the probability density of an in-gap eigenstate near zero energy [marked by the red arrow in Fig.~\ref{fig5}(c)] for a finite QL sample with square boundary geometry under the OBC. The red dots (blue circles) represent an edge state with spin up (down). It is found that the in-gap states are located at the square edge of the finite QL sample.

It is necessary to calculate the real $Z_2$ invariant beyond the spin Bott index for the class DIII TRI TSC. However, it is hard work to calculate directly the generic $Z_2$ invariant in a quasicrystal. On the other hand, we can indirectly know the value of the $Z_2$ invariant based on the bulk-boundary correspondence. Thus, we can perform numerical calculations of the energy spectrum of the Hamiltonian~(\ref{HDIII}) on the quasicrystalline lattice with a square shape under the OBC and PBC to present helical edge states appearing in the bulk gap. The indirect $Z_2$ invariant can be obtained, and the phase with the helical edge states corresponds to $Z_2=-1$ (the topologically nontrivial phase), while the phase without the edge states corresponds to $Z_2=1$ (the topologically trivial phase). The original $Z_2$ invariant for the generic form of the class DIII TSC Hamiltonian will be investigated in our future work.

\section{Conclusion and discussion}
\label{Conclusion}

In this work, we investigated the topological phase transitions of a class D chiral TSC and a class DIII TRI TSC with Anderson-type disorder in an AB tiling QL. We employed real-space topological invariants, including the Bott index (a $\mathbb{Z}$ index for the class D system) and the spin Bott index (a $\mathbb{Z}_2$ index for the class DIII system), and the two-terminal thermal conductance to determine the topological phases of the two quasicrystalline TSC systems. The class D chiral TSC in the topologically nontrivial phase exhibits chiral MEMs located at the square boundary of a finite QL sample, and the class DIII TRI TSC has helical MEMs. Both the chiral MEMs in the class D TSC system and the helical MEMs in the class DIII TSC system are robust against weak disorder, while they are destroyed when the disorder is strong. More striking is that we discovered a topological phase transition from a topologically trivial phase to a topologically nontrivial phase with chiral MEMs located on the edge of the class D quasicrystalline TSC at finite disorder strength. Similarly, disorder-induced helical MEMs in the class DIII quasicrystalline TSC were also found. We also presented the phase diagrams based on the numerical calculation of the Bott index and the spin Bott index as functions of the disorder strength and the chemical potential, and we showed that the interplay between the model parameters and disorder has an interesting influence on the existence of the topological phases in the quasicrystalline TSC systems.

The theoretical interpretation of the disorder-induced topological phase is that the model parameters are renormalized by the disorder, which is obtained with the effective-medium theory (self-consistent Born approximation method) in crystalline systems \cite{PhysRevLett.103.196805,PhysRevB.93.125133,Qin2016SR,PhysRevB.103.115430,PhysRevB.100.205302}. However, because the translational symmetry is lacking in QLs, the self-consistent Born approximation method is invalid. The disorder-induced chiral and helical MEMs in the AB tiling QL cannot be properly explained by the effective-medium theory. Considering the similarities of the disorder-induced topological phases in crystalline and quasicrystalline lattices, we can conjecture that the origin of the appearance of the disorder-induced MEMs in the AB tiling QL is renormalization of the model parameters, which is caused by disorder.

Furthermore, the subgap Yu-Shiba-Rusinov bound states, which are induced by magnetic impurity atoms in a superconductor, can be employed in forming a TSC \cite{Li_2020YSR}, such as the 1D TSC chain \cite{Nadj-Perge602,Jeon772} and the 2D amorphous TSC \cite{P_yh_nen_2018}. Therefore, we propose the experimental setup of the chiral TSC in the AB tiling QL is that the magnetic atoms, which are located at the vertices of the QL, are placed on a superconducting surface, while the TRI TSC in the AB tiling QL is formed by a heterostructure, which consists of a layer of atoms being placed between two superconductors, where the superconductors have a $\pi$ phase difference \cite{PhysRevLett.100.096407,PhysRevB.79.161408}.

\section*{Acknowledgments}
B.Z. was supported by the NSFC (under Grant No. 12074107) and the program of outstanding young and middle-aged scientific and technological innovation team of colleges and universities in Hubei Province (under Grant No. T2020001). D.-H.X. was supported by the NSFC (under Grant No. 12074108). D.-H.X. also acknowledges the financial support of the Chutian Scholars Program in Hubei Province.

\bibliographystyle{apsrev4-1-etal-title_6authors}

\end{document}